\documentclass[12pt,a4paper]{article}
\usepackage{amsmath,epsfig,amssymb}
\usepackage{amscd,latexsym}
\usepackage{color}
\usepackage[bf,footnotesize]{caption}
\usepackage[rmargin=2.5cm,lmargin=2.5cm]{geometry}
\usepackage[dvips,a4paper]{hyperref}

\date{}

\def\lsi{\raise0.3ex\hbox{$<$\kern-0.75em\raise-1.1ex\hbox{$\sim$}}}
\def\gsi{\raise0.3ex\hbox{$>$\kern-0.75em\raise-1.1ex\hbox{$\sim$}}}

\newcommand{\mps}{m_\mathrm{PS}}
\newcommand{\fchips}{f_{\chi}^{\mathrm{PS}}}
\newcommand{\fps}{f_\mathrm{PS}}
\newcommand{\mv}{m_\mathrm{V}}

\newcommand{\mchipcac}{m^\mathrm{PCAC}_\chi}
\newcommand{\rAP}{r_{\hspace{-.4ex}A\hspace{-.1ex}P}}

\graphicspath{{plots/}}

\begin{document}

\begin{titlepage}

\title{
  {\vspace{-2cm} \normalsize
  \hfill \parbox{40mm}{DESY 05-096\\
                       SFB/CPP-05-23\\
                       MS-TP-05-15\\June 2005}}\\[10mm]
Lattice Spacing Dependence \\  of the First Order Phase Transition \\ 
for Dynamical Twisted Mass Fermions}  

\author{F. Farchioni$^{\, 1}$,
K. Jansen$^{\, 2}$,
I. Montvay$^{\, 3}$,
E.E. Scholz$^{\, 3}$,
L. Scorzato$^{\, 4}$,\\
A. Shindler$^{\, 2}$,
N. Ukita$^{\, 3}$,
C. Urbach$^{\, 2,\, 5}$,
U. Wenger$^{\, 2}$ and I. Wetzorke$^{\, 2}$  
\vspace*{0.6cm} \\
{\small $^{1}$
Institut f\"{u}r Theoretische Physik, Universit\"{a}t M{\"u}nster,}\vspace*{-.2cm}\\
{\small Wilhelm-Klemm-Str. 9, 48149 M{\"u}nster, Germany}
\vspace*{2mm}\\
{\small $^{2}$   NIC,
Platanenallee 6, 15738 Zeuthen, Germany}
\vspace*{2mm}\\
{\small $^{3}$   DESY, Notkestr. 85, 22607 Hamburg, Germany}
\vspace*{2mm}\\
{\small $^{4}$ Institut f\"{u}r Physik, Humboldt Universit\"{a}t zu Berlin,}\vspace*{-.2cm}\\
{\small Newtonstr.\ 15, 12489 Berlin, Germany}
\vspace*{2mm}\\
{\small $^{5}$
Institut f\"{u}r Theoretische Physik, Freie Universit\"{a}t Berlin,}\vspace*{-.2cm}\\
{\small Arnimallee 14, 14195 Berlin, Germany} }

\maketitle
\thispagestyle{empty}
\begin{abstract}
Lattice QCD with Wilson fermions generically shows the phenomenon of a
first order phase transition. 
We study the phase structure of lattice QCD using Wilson twisted mass
fermions and the Wilson plaquette gauge action in a range of
$\beta$ values where such a first order phase transition is observed. In
particular, we investigate the dependence of the  
first order phase transition on the value of the lattice
spacing. Using only data in one phase and neglecting possible problems
arising from the phase transition we are able to perform a first
scaling test for physical quantities using this action.
\vspace{0.75cm}
%%%{\it PACS:}  ; ; \\
%%%{\it Keywords:}  ; .
\end{abstract}

\end{titlepage}
\renewcommand{\baselinestretch}{1.1}\normalsize

\section{Introduction}

Understanding the phase structure of lattice QCD 
is an important pre-requisite before starting large scale 
simulations. Indeed, our collaboration found  that when working at
lattice spacings of about $0.15\ \mathrm{fm}$ there can be strong first order
phase transitions at small quark masses,  at least when a
combination of  Wilson plaquette action and Wilson fermions is used 
\cite{Farchioni:2004us,Farchioni:2004ma}.  The phenomenon appears
also when a small twisted mass term is switched on. 
This has serious consequences, since in such a scenario the pion mass
$\mps$ cannot be made arbitrarily small but assumes a minimal value,
$\mps^\mathrm{min}$, which may be about $500\ \mathrm{MeV}$ and hence
it becomes  impossible to work close to the physical value of the pion
mass.

The presence of the first order phase transition for pure Wilson 
fermions is in accordance with predictions from chiral perturbation
theory \cite{Sharpe:1998xm}, which have been extended later to the case
of adding a twisted mass
\cite{Munster:2004am,Scorzato:2004da,Sharpe:2004ny,Sharpe:2004ps,Aoki:2004ta}.
Let us, for completeness, also mention that for values of the lattice
spacing much coarser than $a=0.15\ \mathrm{fm}$ the first order phase
transition turns into a second order one from the normal QCD phase to the
so-called Aoki phase \cite{Aoki:1984qi,Ilgenfritz:2003gw,Sternbeck:2003gy}. The
generic phase structure of lattice QCD according to our present
understanding is discussed and illustrated in
refs.~\cite{Farchioni:2004us,Farchioni:2004ma,Farchioni:2004fs}.

In refs.~\cite{Farchioni:2004us,Farchioni:2004ma} we have studied only 
one value of the inverse gauge coupling $\beta=6/g_0^2$ in order to 
demonstrate the existence of the first order phase transition, 
leaving the question of the $\beta$ dependence open. Since lattice chiral
perturbation theory  predicts a weakening of the first order phase
transition towards the continuum limit, it is interesting to check
this prediction and, in particular, to investigate quantitatively  
how fast the transition weakens when the continuum limit is
approached. The answer to the latter question will naturally depend on
the choice of the actions that are used for the gauge and the fermion
fields.

In this paper we will present results using Wilson twisted mass 
fermions and the Wilson plaquette gauge action for three values of 
$\beta$. At each of these $\beta$ values we have performed simulations at a
number of quark masses on both sides of the first order phase transition. 
This allows to study the $\beta$ dependence of the phase transition itself
and, in addition, the lattice spacing dependence of physical
observables computed separately in the two phases. We have performed
such a scaling test for the pion mass,
the pion decay constant and the ratio of the pion to the vector
meson mass. For a scaling test of Wilson twisted mass fermions and
other recent results in the quenched approximation see
refs.~\cite{Jansen:2003ir,Abdel-Rehim:2005gz,Jansen:2005gf}.

\section{Wilson twisted mass fermions}

In this paper we will work with Wilson twisted mass fermions
\cite{Frezzotti:2000nk} that can be arranged to be $O(a)$ improved
without employing specific improvement terms
\cite{Frezzotti:2003ni}. The Wilson tmQCD action in the twisted basis
can be written as 
\begin{equation}
  \label{tmaction}
  S[U,\chi,\bar\chi] = a^4 \sum_x \bar\chi(x) ( D_W + m_0 + i \mu
\gamma_5\tau_3 ) \chi(x)\; ,
\end{equation}
where the Wilson-Dirac operator $D_{\rm W}$ is given by
\begin{equation}
  D_{\rm W} = \sum_{\mu=0}^3 \frac{1}{2} 
  [ \gamma_\mu(\nabla_\mu^* + \nabla_\mu) - a \nabla_\mu^*\nabla_\mu]
  \label{Dw}
\end{equation}
and $\nabla_\mu$ and $\nabla_\mu^*$ denote the usual forward
and backward derivatives and the Wilson parameter $r$ was set to $1$. 

The situation of full twist and hence automatic $O(a)$ improvement
arises when $m_0$ in eq.~(\ref{tmaction}) is tuned towards a critical
bare quark mass $m_{\mathrm{crit}}$. We use for our simulations the
hopping representation of the Wilson-Dirac operator with
$\kappa=(2am_0+8)^{-1}$.

We extract the pseudo scalar mass $\mps$ and the vector meson mass 
$m_\mathrm{V}$ from the usual correlation functions:
\begin{equation}
  \begin{split}
    C_{PP} (x_0) &= a^3\sum_{\mathbf x} \langle P^+(x)P^-(0)\rangle\ , \\
    C_{VV} (x_0) &= \frac{a^3}{3}\sum_{k=1}^3\sum_{\mathbf x} \langle V_k^+
    (x)V_k^- (0)\rangle\ ,\\
  \end{split}
\end{equation}
where we consider the local bilinears 
$P^\pm =\bar\chi\gamma_5\frac{\tau^\pm}{2}\chi$ and
$V_\mu^\pm =\bar\chi\gamma_\mu\frac{\tau^\pm}{2}\chi$.
Here we used $\tau^\pm=(\tau_1\pm i\tau_2)$ with $\tau_{1,2}$ the first two Pauli
matrices. Similarly one can define the correlation function
$C_{AP}$ with the local bilinear $A_\mu^\pm = \bar\chi\gamma_\mu\gamma_5\frac{\tau^\pm}{2}\chi$.

The bare pseudo scalar decay constant $\fchips$ in the twisted basis can
be obtained from (cf. \cite{Baxter:1993bv,Farchioni:2002vn}) 
\begin{equation}
  \label{fpi}
  \fchips = \mps^{-1}\,\rAP\, \langle0|P^+(0)|\pi\rangle \, ,
\end{equation}
where the ratio
\begin{equation}
  \label{obs:8}
  \rAP = \frac{\langle0|A_0^+(0)|\pi\rangle}{\langle0|P^+(0)|\pi\rangle}
\end{equation}
can be extracted from the asymptotic behavior of 
\begin{equation}
  \label{obs:9}
  \frac{C_{AP}(x_0)}{C_{PP}(x_0)} = \rAP\,\tanh[\mps\,(T/2-x_0)]\, .
\end{equation}
The bare PCAC quark mass $\mchipcac$ in the twisted basis can then be
computed from the ratio 
\begin{equation}
  \label{mPCAC}
  \mchipcac = \frac{\fchips}{2\,\langle0|P^+(0)|\pi\rangle}\,\mps^2\, .
\end{equation}
The sign of $\mchipcac$ and $\fchips$ is determined by the sign of
$\rAP$ and therefore, the corresponding values can be negative. One
has to keep in mind that $\mchipcac$ and $\fchips$, since measured in
the twisted basis, do not correspond to the physical quark mass and the
physical pseudo scalar decay constant, respectively. While the quark
mass is given by a combination of the (renormalized) values of
$\mchipcac$ and $\mu$, the pseudo scalar decay constant can be computed
by the help of $\fchips$ and the twist angle, as long as $\fchips\neq0$
and the value of the twist angle is different from $\pi/2$.

Note that the purpose of the present paper is \emph{not} to work 
at full twist nor to extract physical quantities, but rather to study
the lattice spacing dependence of the first order phase transition.
For the same reason, we also do not address the question of the choice of 
the critical quark mass in order to stay at full twist here, 
see refs.~\cite{Jansen:2005gf,Abdel-Rehim:2005gz} for recent quenched 
simulations addressing this point.

\section{The phase transition as a function of the lattice spacing}

\begin{table}[t]
  \centering
  \begin{tabular*}{.5\textwidth}{@{\extracolsep{\fill}}cccc}
    \hline\hline
    $\beta$ & $\Bigl.L^3\times T\Bigr.$ & $a\mu$ & $a~[\mathrm{fm}]$\\
    \hline\hline
    $5.1$ & $12^3\times 24$ & $0.013$ & $0.200(2)$\\
    $5.2$ & $12^3\times 24$ & $0.010$ & $0.160(4)$\\
    $5.3$ & $16^3\times 32$ & $0.008$ & $0.138(8)$\\
    \hline\hline
  \end{tabular*}
  \caption[Simulation points for Wilson plaquette gauge action.]
  {Simulation points for Wilson plaquette gauge action. For the three
    values of $\beta$ we give the lattice extent, the value for $a\mu$ and
    the value of the lattice spacing in $\mathrm{fm}$, determined
    using $r_0=0.5\ \mathrm{fm}$ at the reference point (see text), where
    $(r_0\mps)^2=1.5$.}
  \label{tab:simphase}
\end{table}

In order to study the lattice spacing dependence of the phase
transition we have chosen three values of $\beta$: $\beta=5.1$, $\beta=5.2$ and
$\beta=5.3$. We scaled the volumes and the values of $\mu$ such that the
physical volume is larger than $2\ \mathrm{fm}$, roughly constant and
that $r_0\mu\approx0.03$, where $r_0$ is the Sommer scale \cite{Sommer:1993ce} 
fixed to be $r_0=0.5\ \mathrm{fm}$. Note that the value of
$r_0/a$ depends on the value of the quark mass and therefore we had to
choose a reference value for $r_0/a$ as will be explained below. The
parameters are summarized in table \ref{tab:simphase}.

\begin{figure}[t]
  \centering
  \includegraphics[width=.8\linewidth]
  {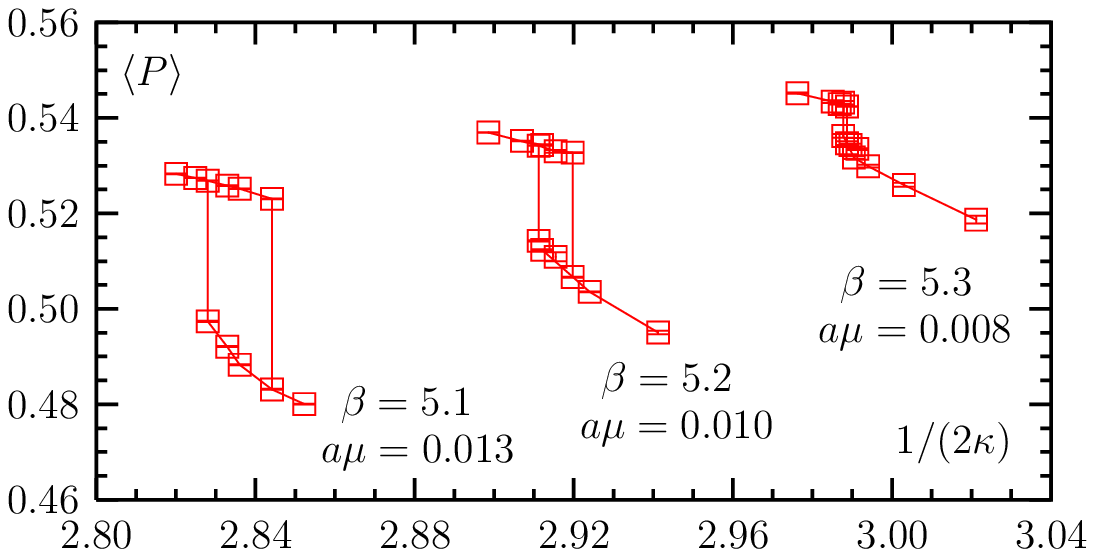}
  \caption{The plaquette expectation value $\langle P\rangle$ as a function of
    $1/(2\kappa)$ at the three values of $\beta$ we have simulated. 
    We also indicate the values of $a\mu$ which are scaled with $\beta$
    such that $r_0\mu$ is roughly constant. The lines just connect the data
    points and only serve to guide the eye. For this study we have more 
    simulations points than in the next figures and than in 
    tables \ref{tab:par5.1}, \ref{tab:par5.2} and \ref{tab:par5.3}.}
  \label{fig:UP}
\end{figure}

In practice it turned out that a very direct way of detecting the
presence of a first order phase transition in lattice QCD is to
monitor the behavior of the plaquette expectation value $\langle P\rangle$,
e.g. as a function of $\kappa$ for fixed twisted mass parameter $\mu$. In
such a situation, starting at identical parameter values from ``hot''
(random) or ``cold'' (ordered) configurations, $\langle P\rangle$ can assume
different, co-existing values. In fig.~\ref{fig:UP} we show $\langle P\rangle$ as
a function of $1/(2\kappa)$ for the three values of $\beta$. The picture is typical
for the behavior of a first order phase transition with meta-stable
branches, one with a low value of $\langle P\rangle$ and one with a high value of
$\langle P\rangle$. We will denote in the following these branches as high (``H'')
and low (``L'') plaquette phases, respectively. 

The $\beta$-dependence shows that the gap in the plaquette expectation
value $\Delta P$ decreases substantially when moving from $\beta=5.1$ ($a\approx
0.20\ \mathrm{fm}$) to $\beta=5.3$ ($a\approx0.12\ \mathrm{fm}$), which is
presumably due to the mixing with the chiral condensate as discussed
in \cite{Farchioni:2004us}. One possible
definition for the quantity $\Delta P$ is the difference between low and
high phase plaquette expectation value at the smallest value of $\kappa$
where a meta-stability occurs.

Let us remark that the first order phase transition exists also in the
continuum limit at zero quark mass where the scalar condensate has a
jump as a consequence of spontaneous chiral symmetry breaking. This
means, of course, that in the continuum limit the phase transition
occurs only for $\mu=0$. 

We give our simulation parameters, the statistics of the Monte Carlo
runs and the results for $a\mps$, $a\fchips$, $a\mchipcac$ and $r_0/a$ in
tables \ref{tab:par5.1}, \ref{tab:par5.2} and \ref{tab:par5.3}.

\begin{figure}[t]
  \centering
  \includegraphics[width=.68\linewidth, height=6cm]
  {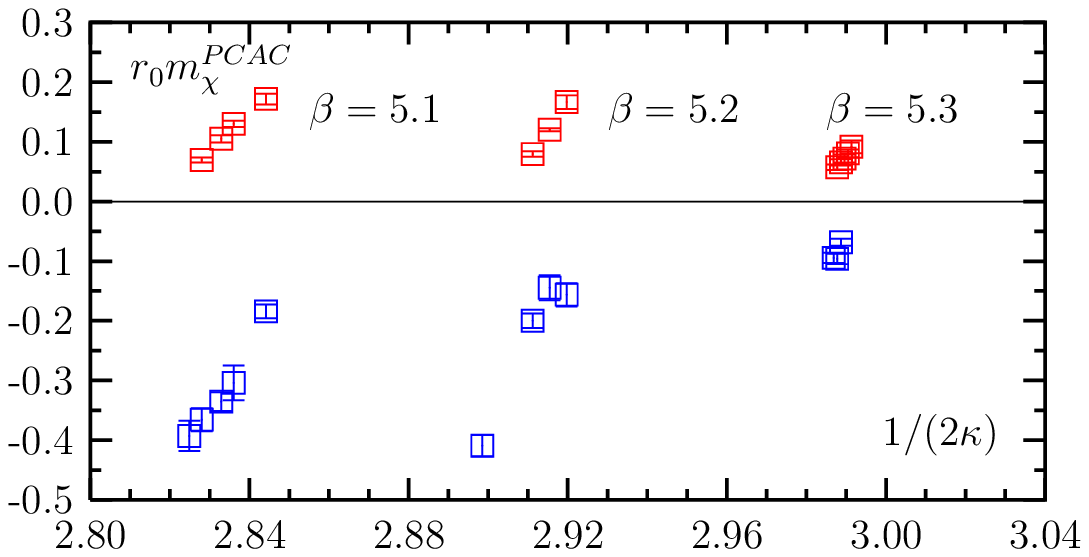}
  \includegraphics[width=.28\linewidth, height=6cm]
  {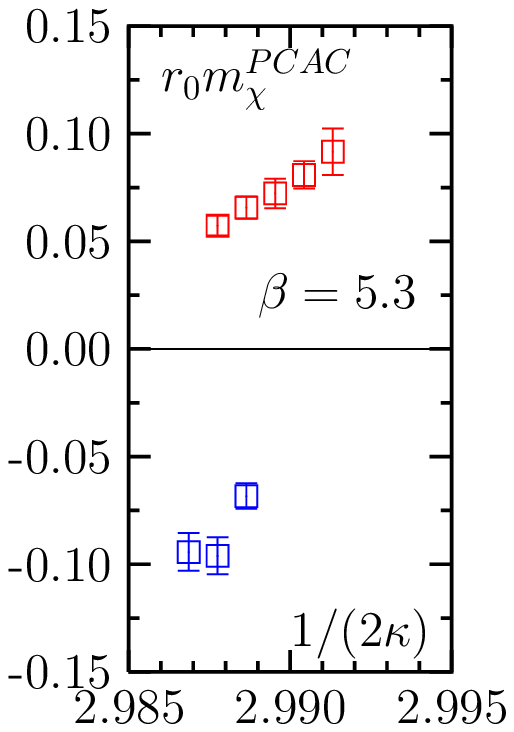}
  \caption{In the graph on the left the PCAC quark mass is plotted as a
    function of $1/(2\kappa)$ at the three values of $\beta$ we have
    simulated. Positive values correspond to the low plaquette phase
    while negative values correspond to the high plaquette phase. The
    statistical errors are on this scale for most of the points
    smaller than the symbols. In the right plot we give a closeup of
    the $\beta=5.3$ results.} 
  \label{fig:mpcac}
\end{figure}

The meta-stability phenomenon observed in $\langle P\rangle$ can also be
seen in fermionic quantities. As an example, we show in 
fig.~\ref{fig:mpcac} the values of the PCAC quark mass 
as obtained in the branches with high and low plaquette expectation
values of fig.~\ref{fig:UP} for the three values of $\beta$. Again we
observe that with increasing $\beta$ the gap between positive  
(low plaquette phase) and negative (high plaquette phase) quark masses 
shrinks. Also, the meta-stability region in $1/(2\kappa)$ gets much narrower
with increasing $\beta$. 

\begin{figure}[t]
  \centering
  \includegraphics[width=.8\linewidth]
  {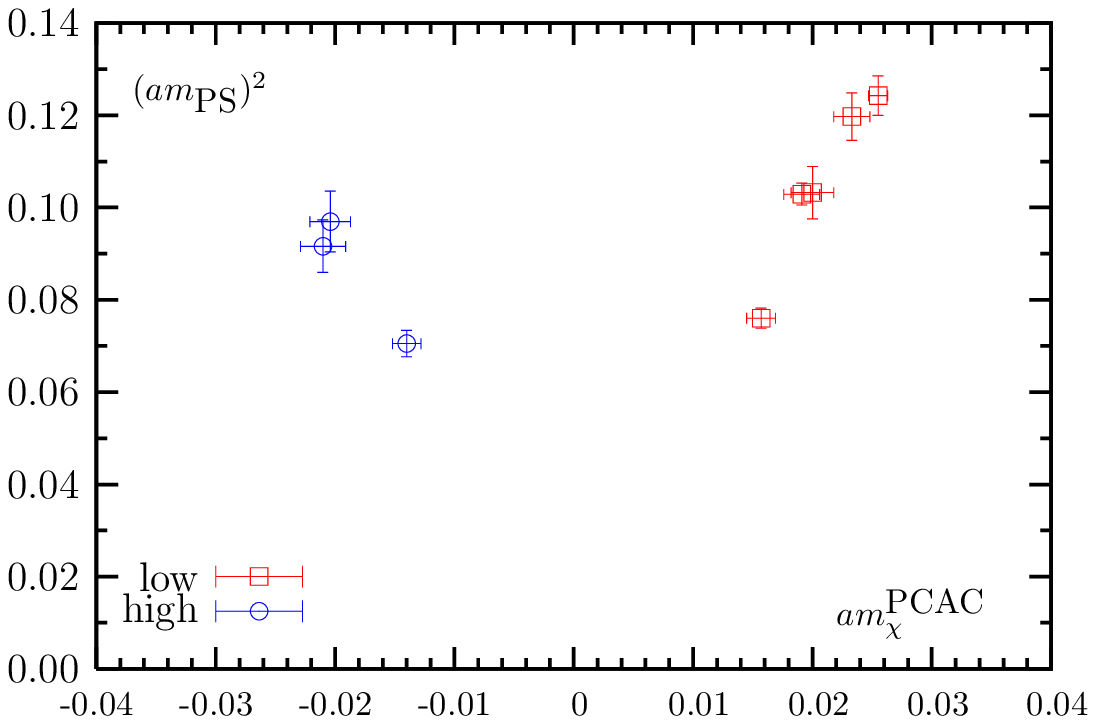}
  \caption{The squared pion mass as a function of the PCAC quark mass
    at $\beta=5.3$.}
  \label{fig:mpion}
\end{figure}

The effects of the first order phase transition can also be seen in
the pion mass and the value of the force parameter $r_0$. We plot in 
fig.~\ref{fig:mpion} an example of the pion mass as a function of the
PCAC quark mass at $\beta=5.3$. The most intriguing observation here is
that due to the presence of the first order phase transition, the pion
mass, say for positive quark masses, does not go to zero but rather
reaches a minimal value, and jumps then to the phase with
negative quark mass. This is, of course, just another manifestation of
the jump in the PCAC quark mass in fig.~\ref{fig:mpcac}.

\begin{figure}[t]
  \centering
  \includegraphics[width=.8\linewidth]
  {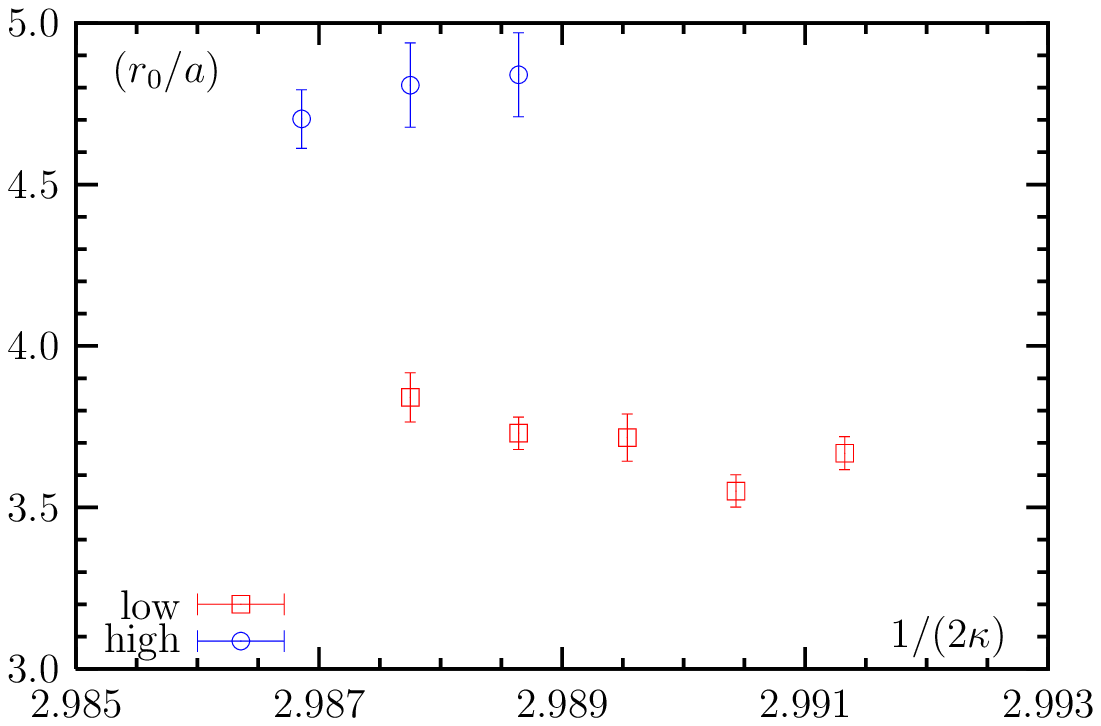}
  \caption{$r_0/a$ as a function of $1/(2\kappa)$ at $\beta=5.3$.}
  \label{fig:r0}
\end{figure}

In fig.~\ref{fig:r0} we also show the values of $r_0/a$ in the low and high 
plaquette phases at $\beta=5.3$. Note that the values of $r_0/a$ are quite
different when determined in the low and the high plaquette phases,
which is a generic feature also for other values of $\beta$ and even for
different gauge actions, see ref.~\cite{Farchioni:2004fs}.

An interesting question is, at which value of the lattice spacing $a$ 
the minimal pion mass $\mps^\mathrm{min}$ assumes a value of, say, 
$300\ \mathrm{MeV}$ where contact to chiral perturbation could be
established.

The pion mass assumes two different values for a fixed quark mass,
once this quark mass lies inside the meta-stability region. These two
values for the pion masses correspond to the two phases that for a
certain interval of quark masses co-exist. The precise determination
of the meta-stability region is, of course, very difficult. We can,
however, give an interval in $\kappa$, $[\kappa_1, \kappa_2]$, that can be read from
tables \ref{tab:par5.1}, \ref{tab:par5.2} and \ref{tab:par5.3} for the
three different $\beta$ values, where meta-stabilities occur in our
simulation. In the following, we will mainly concentrate on the low
plaquette phase since this is the natural choice for studying lattice
QCD. Being interested only in the low plaquette phase we determine
then a lower bound for the minimal pion mass as computed at the lower
end of this interval, i.e. $\kappa_1$, in the low plaquette phase. We give
in table~\ref{tab:mini} the values of the minimal pion masses in the
low plaquette phase in physical units. In addition, we provide the
value for the gap in the plaquette expectation value $\Delta P$.

\begin{table}[t]
  \centering
  \begin{tabular*}{.4\textwidth}{@{\extracolsep{\fill}}ccc}
    \hline\hline
    $\Bigl.\beta\Bigr.$ & $\mps^\mathrm{min}[\mathrm{MeV}]$ & $\Delta P$ \\  
    \hline\hline
    $5.1$ & $\gtrsim 600$ & $0.0399(1)$ \\
    $5.2$ & $\gtrsim 630$ & $0.0261(1)$ \\
    $5.3$ & $\gtrsim 470$ & $0.0077(4)$ \\
    \hline\hline
  \end{tabular*}
  \caption[Lower bound for the minimal pion mass in the low plaquette
  phase and $\Delta P$ for the three $\beta$ values.]
  {Minimal pion mass $\mps^\mathrm{min}$ in physical units in the low
    plaquette phase and $\Delta P$ for the three $\beta$ values. To set the
    scale we used $r_0=0.5\ \mathrm{fm}$ and the value of $r_0/a$
    measured for the corresponding simulation point.}
  \label{tab:mini}
\end{table}

In principle, it would be interesting to extrapolate the minimal 
pion mass and the gap in $\langle P\rangle$ as a function of the lattice
spacing. However, our present data do not allow for a reliable and
safe extrapolation. First of all, the determination of the minimal
pion mass has a large ambiguity in itself since we do not know exactly
for which value of the quark mass the meta-stability will disappear.
A substantially larger statistics would be necessary to answer this
question and to check whether tunneling from one phase to the other  
occurs. Second, the only three values of $\beta$ we have used give a too
short lever arm to perform a trustworthy extrapolation. And, third,
the values of $r_0/a$ are very different in the two phases, as can be
seen in fig.~\ref{fig:r0}, which makes it particularly difficult to
follow the gap in $\langle P\rangle$ as a function of $a/r_0$.

Nevertheless, an estimate on a more qualitative level yields a value 
of the lattice spacing of $a\sim 0.07\ \mathrm{fm} - 0.1\
\mathrm{fm}$ where simulations with pion masses of about $300\
\mathrm{MeV}$ can be performed without being affected by the first
order phase transition. 

\section{Lattice spacing dependence of \\ physical observables} 

Although the present simulations are not at full twist, the fact that 
we have results at three values of $\beta$ with roughly constant $r_0\mu$
allows us to check for the size of lattice artifacts. In order to
perform such an investigation it is advantageous to express physical
quantities in dimensionless variables. To this end,  let us first
define a reference pion mass through $(r_0 \mps)^2=1.5$. We have
chosen this particular value in order to be able to interpolate for
the values of $\beta=5.1$ and $\beta=5.3$, and to perform only a short
extrapolation for $\beta=5.2$ to this point. 

At the aforementioned reference pion mass, a corresponding reference
value of $r_0/a$ and a reference quark mass can be determined, the
latter leading to a variable $\sigma$,   
\begin{equation}
  \sigma=\frac{\mchipcac}{\bigl.\mchipcac\bigr|_\mathrm{ref}}\; .
  \label{eq:sigma}
\end{equation}
Similarly, we can define ratios for a quantity $O$,
\begin{equation}
  R_O=\frac{O}{\bigl.O\bigr|_\mathrm{ref}}
  \label{eq:RO}
\end{equation}
where $O|_\mathrm{ref}$ is the quantity as determined at 
the reference pion mass. The values for several quantities at the
reference point can be found in table \ref{tab:ref}.

In order to determine the reference values for $\mchipcac$, $\fchips$
and $r_0$, in a first step we interpolated $\mchipcac$ linearly as a
function of $(r_0\mps)^2$ to the point where $(r_0\mps)^2=1.5$ and
extracted the reference value for $\mchipcac$. Then we determined the
reference values for $\fchips$ and $r_0$ by quadratically
interpolating the data as a function of $\mchipcac$ to the reference
value of $\mchipcac$. We repeated the latter step with a linear
interpolation finding agreement within the errors.
The fits to the data have been performed with
the ROOT and MINUIT packages from CERN (cf.~\cite{root,minuit}),
taking the errors on both axis into account. We remark that for the
quantity $r_0\mps$ we have neglected the correlation of the data
between $r_0/a$ and $a\mps$.

\begin{table}[tb]
  \centering
  \begin{tabular*}{.7\textwidth}{@{\extracolsep{\fill}}lccc}
    \hline\hline
    $\Bigl.\beta\Bigr.$ & $\Bigl.a\mchipcac\bigr|_\mathrm{ref}$ &
    $\bigl.a\fchips\bigr|_\mathrm{ref}$ &
    $\bigl.(r_0/a)\bigr|_\mathrm{ref}$\\   
    \hline\hline
    $5.1$ & $0.035(2)$ & $0.195(06)$ & $2.497(29)$\\
    $5.2$ & $0.025(4)$ & $0.139(15)$ & $3.124(85)$\\
    $5.3$ & $0.022(1)$ & $0.122(07)$ & $3.628(60)$\\
    \hline\hline
  \end{tabular*}
  \caption{Reference values for $a\mchipcac$, $a\fchips$ and $r_0/a$. The
  reference point is chosen such that $(r_0
  \mps)^2=1.5$. The errors include the interpolation
  errors.}
  \label{tab:ref}
\end{table}

\begin{figure}[t]
  \centering
  \includegraphics[width=0.8\linewidth]
  {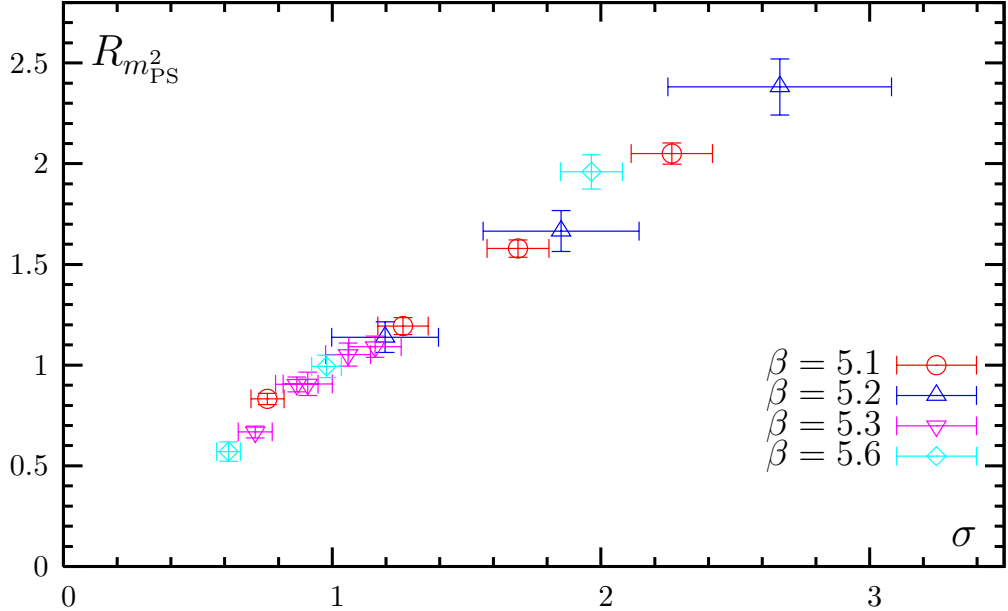}
  \caption{The dependence of the pion mass on the quark mass, 
    i.e.  $R_{\mps^2}$ as a function of $\sigma$. 
    Besides the data of this work, we added in the plot also results 
    from Wilson fermion simulations at $\beta=5.6$ \cite{Urbach:2005ji}
    which were obtained at $\mu=0$.}
  \label{fig:Rmpi}
\end{figure}

For a given observable $O$, $R_O$ is a universal function of $\sigma$ for
fixed value of $\mu$ in physical units that allows for a direct
comparison of results obtained at different values of $\beta$ and,
in principle, even for different actions. Deviations of results at
different $\beta$ values provide then a direct measure of scaling
violations. In fig.~\ref{fig:Rmpi} we show $R_{\mps^2}$  as a function
of $\sigma$. Note that for the scaling analysis we take the data in the low
plaquette phase only since this corresponds to the standard lattice
QCD situation. We also remark that some of the points taken in this
analysis might be meta-stable. Nevertheless, we assume here that these
data can serve for checking scaling violations. Besides the data from
the present work, we added also results from simulations at $\beta=5.6$
\cite{Urbach:2005ji}, which were obtained, however, at vanishing
twisted mass parameter $\mu=0$.

\begin{figure}[htb]
  \centering
  \includegraphics[width=.8\linewidth]
  {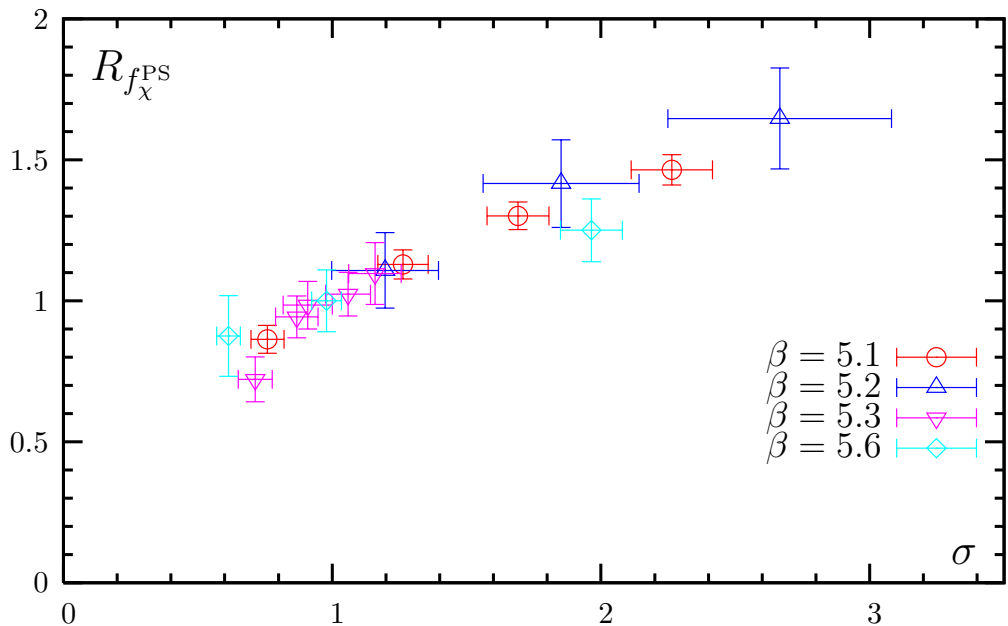}
  \caption{The ratio $R_{\fchips}$ for the pseudo scalar decay
    constant as a function of $\sigma$. We also added results from Wilson
    fermion simulations for $R_{\fps}$ at $\beta=5.6$ \cite{Urbach:2005ji}
    obtained with $\mu=0$.}
  \label{fig:Rfpi}
\end{figure}

\begin{figure}[t]
  \centering
  \includegraphics[width=.8\linewidth]
  {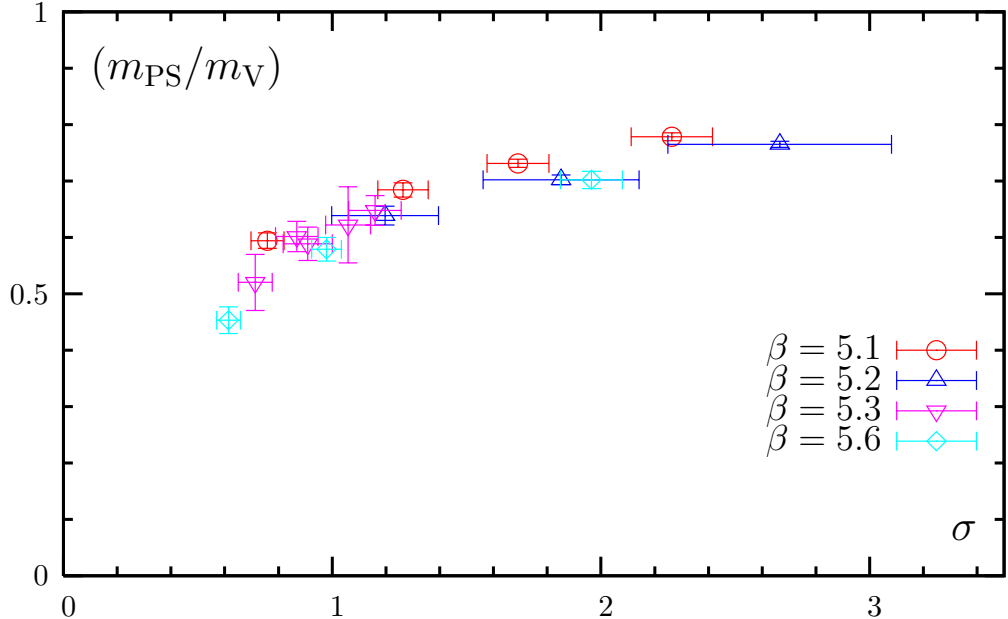}
  \caption{The ratio $\mps/ m_\mathrm{V}$ as a function of $\sigma$. Again,
  we also added results from Wilson fermion simulations at $\beta=5.6$
  \cite{Urbach:2005ji}.}
  \label{fig:Rmpimv}
\end{figure}

A rather amazing consequence of fig.~\ref{fig:Rmpi} is that, despite
the fact that we are using coarse lattices, we cannot detect any
scaling violation, at least within the (large) statistical errors of
our data. Even more, the results of our present simulations at small
values of $\beta$ agree with results from simulations with pure Wilson
fermions at $\beta=5.6$ setting $\mu=0$. The same observation is made for
$R_{\fchips}$, see fig.~\ref{fig:Rfpi} and the ratio $\mps/m_\mathrm{V}$,
see fig.~\ref{fig:Rmpimv}. These results indicate that the lattice
artifacts and the effect of a non-vanishing twisted mass parameter $\mu$
are surprisingly small. We remark here that in the case of the ratios like
$R_{\mps^2}$ and $R_{\fchips}$ one could have cancellation of mass independent 
cutoff effects.
One has also to have in mind that, due to the presence
of the first order phase transition, the simulated pion masses are
still larger than $500\ \mathrm{MeV}$. Whether our findings also hold
when one is approaching the chiral limit is certainly an interesting
but open question. However, in a set-up with Wilson twisted
mass fermions and Wilson plaquette gauge action this question cannot
be answered at these values of the lattice spacing.

\section{Conclusions}

In this paper we have investigated dynamical Wilson twisted mass fermions
employing the Wilson plaquette gauge action. We have performed
simulations at three values of $\beta=5.1, 5.2, 5.3$, corresponding to
values of the lattice spacing of $a \approx 0.20, 0.16, 0.14\ \mathrm{fm}$,
respectively. The non-zero values of the twisted mass parameter $\mu$ were
chosen such that $r_0\mu\approx0.03$ for all of the three $\beta$ values. At
these rather coarse lattice spacings we find clear signals of first
order phase transitions that manifest themselves in a meta-stable
behavior of the plaquette expectation value and fermionic quantities,
such as the PCAC quark mass and the pion mass.

We clearly observe that the gaps in quantities sensitive to the phase
transition, such as the plaquette expectation value and the PCAC quark
mass decrease substantially when $\beta$ is increased. Unfortunately, with our
present set of simulations, we are not able to quantitatively locate
the value of the lattice spacing, where the effects of the first order
phase transition becomes negligible and where a minimal pion mass of, say, 
$300\ \mathrm{MeV}$ can be reached. 
As an estimate of such a value of the lattice spacing we give a range of   
$a\approx 0.07\ \mathrm{fm} -0.1\ \mathrm{fm}$. Of course, this would mean
that a continuum extrapolation of physical results obtained on
lattices with linear extent of at least $L=2\ \mathrm{fm}$ would be
very demanding, since the starting point for such simulations would
already require large lattices. It is therefore very important to find
alternative actions such that the value of the lattice spacing can be
lowered without running into problems with the first order phase
transition. One candidate for such an action, the DBW2 gauge action,
is  discussed in ref.~\cite{Farchioni:2004fs} where it has indeed been
found that modifying the gauge action alone can substantially reduce
the strength  of the first order phase transition. We are presently
investigating another possibility, the tree-level Symanzik improved
gauge action \cite{Weisz:1982zw,Weisz:1983bn}.

Despite the problems arising from the presence of a first order phase 
transition, we performed a scaling analysis for the pion mass, the 
pion decay constant and the ratio $\mps/ \mv$ for the data obtained 
at the three values of $\beta$ where we performed simulations. To this end, 
we only analyzed data from the low plaquette phase, since this is the
natural choice for QCD simulations. 

By defining a reference pion mass at $(r_0 \mps)^2=1.5$, 
we computed the ratio of $\mps$ and $\fchips$ to the corresponding
reference values as a function of the PCAC quark mass, again measured
with respect to the corresponding reference quark mass. We find that
for these ratios the scaling violations are remarkably small and
cannot be detected with the present precision of our data. Even more,
when adding data from simulations of Wilson fermions with $\mu=0$ at
$\beta=5.6$, then these data fall on the same scaling curve as our
results on much coarser lattices and with twisted mass parameter
switched on. This indicates that not only the lattice artifacts 
but also the effect of switching on a twisted mass of the order
of $r_0\mu\approx 0.03$ are small, at least for the rather large pion masses
simulated here. 
This finding is surprising since it suggests that continuum
values of physical quantities can be already estimated from
simulations at not too small lattice spacings.
Of course, our scaling results suffer
from the fact that they are obtained using data that might be meta-stable 
as a consequence of the presence of the first order phase
transition. 
Hence, a scaling test with an action that does not lead to significant 
effects of the first order phase transition is mandatory to check the 
results presented in this paper. 

\section{Acknowledgments}
We thank R.~Frezzotti, G.~M\"unster, G.~C.~Rossi and S.~Sharpe for many useful
discussions. The computer centers at NIC/DESY Zeuthen, NIC at
Forschungszentrum J{\"u}lich and HLRN provided the necessary technical
help and computer resources. We are indebted to R.~Hoffmann and
J.~Rolf for leaving us a MatLab program to check our fits.
This work was supported by the DFG Sonderforschungsbereich/Transregio
SFB/TR9-03.

\begin{table}[htbp]
  \centering
  \begin{tabular*}{1.\textwidth}{@{\extracolsep{\fill}}lllllll}
    \hline\hline
    $\kappa$ &  & $\Bigl.N_\mathrm{meas}\Bigr.$ & $a\mps$ & $a\fchips$ & $a\mchipcac$ & $r_0/a$ \\
    \hline\hline
    $0.1758$ & L & $160$ & $0.7015(031)$ & $+0.2856(60)$ & $+0.0799(12)$ & $2.178(8)(4)(20) $ \\
    $0.1763$ & L & $160$ & $0.6155(040)$ & $+0.2538(56)$ & $+0.0597(12)$ & $2.258(8)(0)(8)  $ \\
    $0.1765$ & L & $160$ & $0.5353(068)$ & $+0.2201(76)$ & $+0.0446(16)$ & $2.370(12)(4)(26)$ \\
    $0.1768$ & L & $160$ & $0.4468(051)$ & $+0.1683(82)$ & $+0.0268(13)$ & $2.625(19)(22)(1)$ \\
    \hline
    $0.1758$ & H & $160$ & $0.5323(126)$ & $-0.2065(119)$ & $-0.0496(25)$ & $3.926(26)(12)(10)$ \\
    $0.1763$ & H & $160$ & $0.6771(116)$ & $-0.2351(227)$ & $-0.0777(50)$ & $4.087(56)(4)(0)  $ \\
    $0.1765$ & H & $160$ & $0.7231(111)$ & $-0.2595(232)$ & $-0.0864(26)$ & $4.053(18)(17)(3) $ \\
    $0.1768$ & H & $160$ & $0.7377(119)$ & $-0.2302(136)$ & $-0.0926(38)$ & $4.139(35)(16)(2) $ \\
    $0.1770$ & H & $160$ & $0.7530(189)$ & $-0.2212(189)$ & $-0.0977(59)$ & $4.045(28)(10)(4) $ \\
    \hline\hline
  \end{tabular*}
  \caption[Parameters and physical observables for the simulations with
    $\beta=5.1$.]
  {Parameters and physical observables for the simulations with
    $\beta=5.1$. The lattice size in these runs was set to $12^3\times 24$ and
    the twisted mass parameter to $a\mu=0.013$. We give the values for
    $\kappa$ and the number of measurements $N_\mathrm{meas}$ performed. We
    indicate with ``L'' or ``H'' whether the plaquette expectation 
    assumes a low or a high value. Moreover, we give the values for
    $\mps$, $\fchips$, $\mchipcac$ and $r_0$ in lattice units. For
    $r_0$ we give in addition to the statistical error two systematic
    errors, the first of them coming from possible excited state
    contaminations and the second from the necessary interpolation of
    the force in $r$.}
  \label{tab:par5.1}
\end{table}

\begin{table}[htbp]
  \centering
  \begin{tabular*}{1.\textwidth}{@{\extracolsep{\fill}}lllllll}
    \hline\hline
    $\kappa$ & & $\Bigl. N_\mathrm{meas}\Bigr.$ & $a\mps$ & $a\fchips$ & $a\mchipcac$ & $r_0/a$ \\
    \hline\hline
    $0.17125$ & L & $320$ & $0.6057(025)$ & $+0.2289(35)$ & $+0.0650(08)$ & $2.618(20)(5)(49)$ \\
    $0.17150$ & L & $459$ & $0.5066(050)$ & $+0.1968(38)$ & $+0.0452(08)$ & $2.800(17)(9)(4) $ \\
    $0.17175$ & L & $320$ & $0.4189(071)$ & $+0.1540(84)$ & $+0.0292(17)$ & $3.038(28)(14)(4)$ \\
    \hline
    $0.17125$ & H & $320$ & $0.4173(111)$ & $-0.1571(166)$ & $-0.0352(43)$ & $4.796(63)(65)(15)$ \\
    $0.17150$ & H & $318$ & $0.4220(126)$ & $-0.1566(219)$ & $-0.0349(50)$ & $4.282(61)(16)(0) $ \\
    $0.17175$ & H & $320$ & $0.4985(088)$ & $-0.1770(119)$ & $-0.0494(28)$ & $4.418(23)(23)(0) $ \\
    $0.17250$ & H & $320$ & $0.6462(131)$ & $-0.1974(087)$ & $-0.0874(24)$ & $4.767(51)(7)(3)  $ \\
    \hline\hline
  \end{tabular*}
  \caption[Parameter and physical observables for the simulations with
    $\beta=5.2$.]
    {Parameter and physical observables for the simulations with
    $\beta=5.2$. The lattice size in these runs was set to $12^3\times 24$ and
    the twisted mass parameter to $a\mu=0.01$. See table
    \ref{tab:par5.1} for further explanations.}
  \label{tab:par5.2}
\end{table}

\begin{table}[htbp]
  \centering
  \begin{tabular*}{1.\textwidth}{@{\extracolsep{\fill}}lllllll}
    \hline\hline
    $\kappa$ & & $\Bigl.N_\mathrm{meas}\Bigr.$ & $a\mps$ & $a\fchips$ & $a\mchipcac$ & $r_0/a$ \\ 
    \hline\hline
    $0.16715$ & L & $100$ & $0.3525(061)$ & $+0.1349(111)$ & $+0.0255(18)$ & $3.668(34)(9)(8) $ \\
    $0.16720$ & L & $101$ & $0.3460(075)$ & $+0.1259(063)$ & $+0.0233(15)$ & $3.551(47)(2)(1) $ \\
    $0.16725$ & L & $180$ & $0.3213(088)$ & $+0.1211(078)$ & $+0.0200(18)$ & $3.716(49)(24)(0)$ \\
    $0.16730$ & L & $243$ & $0.3208(037)$ & $+0.1160(063)$ & $+0.0191(15)$ & $3.730(35)(6)(9) $ \\
    $0.16735$ & L & $160$ & $0.2757(040)$ & $+0.0887(084)$ & $+0.0157(12)$ & $3.841(47)(28)(1)$ \\
    \hline

    $0.16730$ & H & $388$ & $0.2656(054)$ & $-0.1037(093)$ & $-0.0140(12)$ & $4.84(10)(2)(1)  $ \\
    $0.16735$ & H & $100$ & $0.3114(106)$ & $-0.1291(128)$ & $-0.0204(17)$ & $4.808(95)(33)(3)$ \\
    $0.16740$ & H & $100$ & $0.3027(094)$ & $-0.1237(137)$ & $-0.0210(19)$ & $4.703(90)(0)(1) $ \\
    \hline\hline
  \end{tabular*}
  \caption[Parameter and physical observables for the simulations with
    $\beta=5.3$.]
    {Parameter and physical observables for the simulations with
    $\beta=5.3$. The lattice size in these runs was set to $16^3\times 32$ and
    the twisted mass parameter to $a\mu=0.008$. For further explanations
    see table \ref{tab:par5.1}.}
  \label{tab:par5.3}
\end{table}

\newpage
\bibliographystyle{h-physrev4}
\bibliography{wilson}

\end{document}